Entrainment and stimulated emission of auto-oscillators in an acoustic cavity.

Richard L Weaver, Oleg I Lobkis[1], and Alexey Yamilov[2]

[1]Department of Physics, University of Illinois, 1110 W Green St, Urbana, IL  61801

[2]Department of Physics, University of Missouri-Rolla, 1870 Miner Circle, Rolla, Missouri 65409

**Abstract**

We report theory, measurements and numerical simulations on nonlinear piezoelectric ultrasonic devices with stable limit cycles.  The devices are shown to exhibit behavior familiar from the theory of coupled auto-oscillators.  Frequency of auto-oscillation is affected by the presence of an acoustic cavity as these spontaneously emitting devices adjust their frequency to the spectrum of the acoustic cavity.  Also, the auto-oscillation is shown to be entrained by an applied field; the oscillator synchronizes to an incident wave at a frequency close to the natural frequency of the limit cycle.  It is further shown that synchronization occurs here with a phase that can, depending on details, correspond to stimulated emission: the power emission from the oscillator is augmented by the incident field.  These behaviors are essential to eventual design of an ultrasonic system that would consist of a number of such devices entrained to their mutual field, a system that would be an analog to a laser.  A prototype *uaser* is constructed.

**I. Introduction**

Laser oscillation is not necessarily a quantum phenomenon. Classical laser designs[1-6] have pedagogic value but no devices have yet been constructed, and the designs have played no role in laser research. Here we report construction of nonlinear classical oscillators in contact with an acoustic cavity and capable of both spontaneous and stimulated emission. We believe systems consisting of such oscillators will have both pedagogic and research value. Because acoustic waves with their longer wavelengths and longer time scales permit probes and controls to a degree not possible in optics, these systems should permit detailed experiments that complement those possible with lasers.

The sine-qua-non of a laser is *stimulated emission*, in which a wave incident upon an excited oscillator is re-emitted with unchanged phase and increased amplitude. It is perhaps not widely appreciated that stimulated emission is a classical phenomenon[1-6]. An excited classical linear oscillator will exhibit stimulated absorption or stimulated emission depending on the phase of the oscillator relative to that of the incident field. Thus an incoherent mixture of excited classical linear oscillators will show no net stimulated emission. If, however, all or most of the oscillators can be induced to have the same frequency and the correct phases, the set will exhibit stimulated emission. Such systems may be termed laser analogs [7,8].

The classical laser designs of Borenstein and Lamb [1,2] and Kobelov *et al*[4], are composed of incoherently excited Duffing oscillators (i.e., stiffness having a cubic nonlinearity). The oscillators emit spontaneously in a trivial fashion. When they go into resonance with each other with the right phase relation, under the influence of their mutual radiation field, they also emit by stimulated emission. These designs are pedagogically stimulating. No attempt has yet been made to build them.

Zavtrak[5,6] has suggested that bubbles or other particles in a fluid could be pumped by an applied coherent harmonic electric or acoustic field. He suggests that the particles would bunch spatially under the influence of their radiation forces, leading to a coherent re-emission in a direction imposed by their radiation field and the modes of their cavity. This analog to a free–electron laser has not been constructed either.

Recent years have seen considerable interest in dynamic synchronization [9-13] in which sets of simple distinct coupled auto-oscillating limit cycles routinely synchronize. The phenomenon occurs in disparate circumstances, including firefly flashes, brain waves, esophagal waves, bridges with crowds of pedestrians, and chemical oscillations. It occurs amongst lasers, Josephson junctions and pendulum clocks. The chief mathematical model for studying the synchronization of large numbers of auto-oscillators is that of Kuramoto, and its generalizations[7-10]. After arguing that the state of a limit cycle oscillator is well represented in terms of its phase $\xi$, and that the phase of each of the oscillators is weakly coupled to the phases of the others[14], one derives a set of N coupled first order ordinary differential equations:[ 9-12]

$$d\xi_n/dt = \omega_n + (A/N) \ \Sigma_m \sin(\xi_m - \xi_n).$$

It has been shown that this model, in the thermodynamic limit N→∞, exhibits a phase transition in which a macroscopic number of oscillators lock to one and other. Sundry generalizations have been discussed, such as time delays and randomness in the couplings. Resemblance between this phase transition and the onset of coherence amongst the atoms of a laser has been noted[8].

In seeking a classical wave analog to a laser we recognize the necessity for the fixing of local oscillators to a common frequency, and thus their coherence. Synchronization of auto-oscillators is one way in which this can be done. But it is also necessary that these oscillators lock to the mutual field with phases that correspond to stimulated emission rather than absorption. As discussed elsewhere[1,2], a random phase in an oscillator, even if it has the same frequency as an incident field, will lead to no particular stimulated emission. Only to the extent that the oscillator and the incident field have the correct phase relation will the energy in the oscillator be transferred efficiently to the wave field. For an excited atom in an electromagnetic field, this phase relation is automatically satisfied. A classical linear oscillator on a mechanical body applies a force with no particular phase relation to an incident field. Nonetheless dynamic synchronization studies[9-13] show that nonlinear classical auto-oscillators can be entrained by their mutual radiation or by an applied field. They tend to synchronize to a common frequency with fixed phase relation. If that phase relation is such that each oscillator emits at a rate greater than it does without an incident field, there is stimulated emission and the system will be a laser analog. [7]

Here we present an electro-mechanical auto-oscillator that could be an element in an ultrasonic laser analog. When placed on an acoustic cavity the oscillator exhibits both spontaneous and stimulated emission. It is pictured in figure 1. A piezoelectric transducer is attached to a regular or irregular elastic body at arbitrary position. Like most ultrasonic transducers it is reciprocal; it both radiates and receives ultrasound. It has electronic impedance that is nominally capacitive with additional small contributions from the mechanics. Our transducer is part of a nonlinear electronic circuit with a limit cycle at a frequency and amplitude that depends on circuit parameters. A second transducer at *y* monitors the acoustic state of the elastic body. An optional third transducer at *z* is driven by a continuous wave at a prescribed frequency and amplitude.

In the next section we propose an analytic model and predict the frequency and rate of emission at *x* in the absence of the source at *z*. We also predict circumstances, with arguments similar to those of [9-14], under which the oscillator at *x* will entrain to an applied field from *z*. We further describe circumstances under which that entrainment to an applied field corresponds, not just to synchronization, but also to stimulated emission. Later sections present results from numerical simulations and observations from experiments. Finally, it is demonstrated that two or three of our piezoelectric auto-oscillators, when entrained to their mutual ultrasonic field as in [9-14], do with stimulated emission and exhibit super radiance.

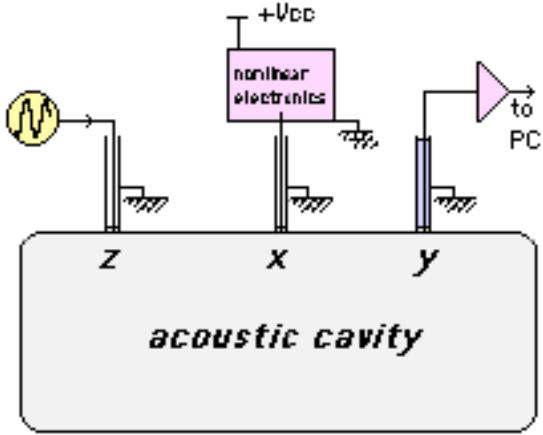

Figure 1. A reverberant acoustic cavity (elastodynamic body), large compared to a wavelength, is driven at position *x* by a piezoelectric transducer that is part of a circuit with an attracting limit-cycle. The acoustic state of the cavity is monitored by a separate receiver at *y*. An optional third transducer at *z* is driven by an applied harmonic signal at prescribed frequency and amplitude.

## II. Entrainment and stimulated emission of piezoelectric auto-oscillators, theory

Here we study the behavior of a linear piezoelectric device that is part of a van-der Pol electronic oscillator [15] and which is in contact with an acoustic cavity. We posit that the transducer is a standard linear two-port network [16], with potential drop *v* and charge *q* on one port, and mechanical force *f* and mechanical displacement *u* on the other. These inputs and outputs are related by a matrix.

$$\begin{pmatrix} v \\ f \end{pmatrix} = \begin{bmatrix} T_{11} & T_{12} \\ T_{21} & T_{22} \end{bmatrix} \begin{pmatrix} q \\ u \end{pmatrix}; \quad \begin{pmatrix} q \\ u \end{pmatrix} = \begin{bmatrix} T_{22}/D & -T_{12}/D \\ -T_{21}/D & T_{11}/D \end{bmatrix} \begin{pmatrix} v \\ f \end{pmatrix} \quad (1a,1b)$$

where $D = T_{11}T_{22} - T_{12}T_{21}$. $T_{11}$ may be interpreted as the inverse of the piezoelectric's capacitance when clamped ($u=0$). $T_{22}/D$ may be interpreted as the capacitance C of the device when there is no force on it ($f=0$). $T_{22}$ is the shortcircuited mechanical stiffness of the device. Reciprocity demands $T_{12} = T_{21}$. All coefficients are in general complex functions of frequency; $T^*(\omega) = T(-\omega)$.

The rate at which the force and electric potential are doing work on the transducer is

$$P = (v \ f) \times \begin{pmatrix} \partial q/\partial t \\ \partial u/\partial t \end{pmatrix} = (q \ u)[T]^T \times \begin{pmatrix} \partial q/\partial t \\ \partial u/\partial t \end{pmatrix}$$

The symbol × represents, in the time domain, a simple multiplication; other juxtapositions represent time-domain convolutions. In the frequency domain these juxtapositions are simple multiplications. If $q$ and $u$ are harmonic: $q = Q\exp(i\omega t) + c.c.$; $u = U\exp(i\omega t) + c.c.$, we may conclude that the time average Power flow into the transducer is

$$P = -i\omega (Q \ U)[T(\omega)^T - T(\omega)^*] \begin{pmatrix} Q^* \\ U^* \end{pmatrix}$$

with $[T(\omega)] = \int \exp(-i\omega t)[T(t)] \, dt = [T(\omega)]^T$. A passive transducer must have $P \geq 0$. We therefore conclude that $[T(\omega)]$'s imaginary part must be positive (negative) semi-definite at positive (negative) frequency.

IIA  A piezoelectric transducer in a nonlinear electronic circuit

We insert the transducer into the circuit of figure 2. Current conservation demands, where $I_L$ is the current in the inductor,

$$I(V - v) = dq/dt + I_L$$

This has a steady solution at $v = 0$, $I_L = I(V)$. Infinitesimal perturbations to the steady solution are governed by

$$-I'(V_c)v = dq/dt + \int v\, dt / L \qquad (2)$$

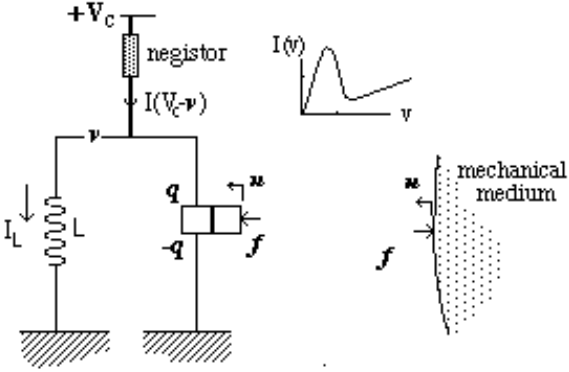

Figure 2] Diagram for a piezoelectric auto oscillator. The current $I$ into the device depends on the potential across the negistor in a nonlinear fashion $I(V_c-v)$ where the function I is characteristic of the negistor. The static supply voltage V is adjusted so that, at static equilibrium, $v=0$ and $dI/dV$ is negative and maximal. At this point current varies with $v$ like $g(v)\, v$, $-g(0)$ being the slope of $I(V)$ at this point. We take g in the form $g = \varepsilon C (1-\beta v^2)$ with $\varepsilon > 0$. Further details of the negistor are described in section IV.

Differentiating (2) with respect to time, and substituting for $d^2q/dt^2$ by (1b) gives:

$$d^2v/dt^2 - \varepsilon(1 - 3\beta v^2)\, dv/dt + (LC)^{-1}v = (T_{12}/T_{22})\, d^2f/dt^2 \qquad (3)$$

This form is useful when the force is prescribed. In particular if $f = 0$, corresponding to the transducer being mechanically free, (3) becomes a van der Pol equation[15,17].

Equation (1b) implies a relation between $f$ $u$ and $v$:

$$f = T_{21}/T_{11}\, v + D/T_{11}\, u \qquad (4)$$

So, if $u$ is prescribed there is an alternate form:

$$d^2v/dt^2 - \varepsilon(1-3\beta v^2)dv/dt + [(LC)^{-1} - \{T_{12}T_{21}/T_{22}T_{11}\}\ d^2/dt^2]v$$
$$= (DT_{12}/T_{22}T_{11})\ d^2u/dt^2 \qquad (5)$$

IIB Spontaneous emission

Equations (3,4,5) are not complete unless we specify *u* or *f,* or a relation between them. A case of primary interest is that of free radiation into an otherwise passive acoustic medium, for which: *u* = - G*f*. (The minus sign arises from the opposite orientation of *f* and *u* with respect to the acoustic medium, see figure 2). Here G is $G_{xx}$, a diagonal element of Greens' function of the acoustic cavity at position *x*. Thus, from (4)

$$[1 + DG/T_{11}]f = T_{21}/T_{11}\ v \qquad (6)$$

and

$$d^2v/dt^2 - \varepsilon(1-3\beta v^2)dv/dt + [(LC)^{-1} - \{T_{12}T_{21}/T_{22}(T_{11}+DG)\}\ d^2/dt^2]v = 0 \qquad (7)$$

Nonlinear dynamical equations such as (7) are challenging. This is especially so if the complicated time-delays and temporal convolutions in T and G are fully expressed. For the present purposes we will presume that G may be evaluated at the frequency of chief interest, that G is only weakly dependent on frequency and can be replaced by a constant. The elements of [T] also vary with frequency but generally in a much smoother fashion, so replacing them by constants is readily justified. Rewriting (7) as a phase oscillator simplifies it [ 14, 7-12 ]. One assumes the oscillator stays on or near its limit cycle, at a frequency Ω, with a slowly varying real amplitude and phase V and ξ:

$$v = V(t)\exp(i\Omega t + i\xi(t)) + c.c. \qquad (8)$$

On assuming $dV/dt \ll \Omega V$; $d\xi/dt \ll \Omega$, and $|\omega - \Omega| \ll \Omega$ we derive

$$[2i\Omega\, dV/dt - \Omega^2 V + \omega^2 V - 2V\Omega d\xi/dt - i\varepsilon\Omega(1 - 6\beta V^2)V + i\mu V]\exp(i\Omega t + i\xi(t)) = 0 \quad (9)$$

where $\omega^2$ is the real part of $[(LC)^{-1} + \{T_{12}T_{21}\Omega^2 / T_{22}[T_{11} + DG]\}]$ and $\mu$ is the imaginary part.

The real and imaginary parts of eqn 9 are:

$$dV/dt - (1/2)\varepsilon(1 - 6\beta V^2)V + (1/2\Omega)\mu V = 0 \qquad (10)$$

and

$$d\xi/dt = (\omega^2 - \Omega^2)/2\Omega \approx (\omega - \Omega) \qquad (11)$$

A steady solution implies $\Omega = \omega$, and $V = [(\varepsilon\Omega - \mu)/6\varepsilon\Omega\beta]^{1/2}$. ( If $\varepsilon\Omega < \mu$, the van der Pol oscillator loses its limit cycle and $V = 0$. )

When the device is unattached to the acoustic cavity, as enforced mathematically by taking $f = 0$, equivalently $G = \infty$, the frequency of the auto oscillation is the real part of $(LC)^{-1/2}$. The frequency is changed by the acoustic cavity to a value $\omega$ that is a solution of the implicit equation $\omega^2 = \text{Re}\,[(LC(\omega))^{-1} + \{T_{12}(\omega)T_{21}(\omega)\omega^2 / T_{22}[T_{11}(\omega) + D(\omega)G(\omega)]\}]^{1/2}$. In a reverberant cavity $G$ can be an irregular function of $\omega$, so solutions may be complicated, or multiple.

The power radiated from the device into the mechanical medium is

$$\Pi = -\boldsymbol{f} \times d\boldsymbol{u}/dt = \boldsymbol{f} \times dG/dt\ \boldsymbol{f} \qquad (12)$$

which is, by 6,

$$\Pi = [1 + D\,G/T_{11}]^{-1} T_{21}/T_{11}\, \boldsymbol{v}\ \times\ dG/dt\ [1 + D\,G/T_{11}]^{-1} T_{21}/T_{11}\, \boldsymbol{v}$$

$$= -2\Omega\,\text{Im}\,G(\Omega)\,|T_{11} + D\,G|^{-2}\,|T_{21}|^2\,V^2 \qquad (13)$$

$$= -2\Omega\,\text{Im}\,G(\Omega)|T_{11} + D\,G|^{-2}\,|T_{21}|^2\,[(\varepsilon\Omega - \mu)/6\varepsilon\Omega\beta]$$

The first equality in (13) arises from a time averaging. ( If $\varepsilon\Omega < \mu$, $\Pi = 0$. ) This is the power of the spontaneous emission.

IIC Entrainment and Stimulated emission

An incident wave field $u^{inc}$ in the acoustic medium can modify the state of the nonlinear oscillator and increase or decrease its emitted power. The oscillator augments the field with its own radiation $-Gf$, so $u = u^{inc} - Gf$. The equation governing the oscillator is still (3), but now with

$$f = T_{21}/T_{11}\, v + D/T_{11}\, [u^{inc} - Gf] \tag{14}$$

so

$$d^2v/dt^2 - \varepsilon(1-3\beta v^2)dv/dt + [(LC)^{-1} - \{T_{12}T_{21}/T_{22}(T_{11}+DG)\}\, d^2/dt^2]v = (DT_{12}/T_{22}(T_{11}+DG))\, d^2 u^{inc}/dt^2 \tag{15}$$

This is identical to (7), but now with a term from the incident field that influences the auto-oscillator. This is a van-der Pol oscillator with harmonic forcing. That such oscillators can be entrained to the frequency of the forcing is well known[2,9-14]. We take the incident field at $x$, $u^{inc}$ to be of the form $U \exp(i\Omega t)$ + c.c. with fixed $\Omega$ and without loss of generality real positive U. The approximations used above now give

$$dV/dt - (1/2)\varepsilon(1-6\beta V^2)V + (1/2\Omega)\mu V = -(U\Omega/2)\operatorname{Im}\{DT_{12}\exp(-i\xi)/T_{22}(T_{11}+DG)\} \tag{16}$$

and

$$d\xi/dt = (\omega - \Omega) + (U/2V)\Omega \operatorname{Re}\{T_{12}\exp(-i\xi)/C(T_{11}+DG)\} \tag{17}$$

This is very similar to Adler's equation[14,9,13].

For sufficiently small detuning $|\omega - \Omega| < \Omega\, (U/2V)\, |T_{12}/C(T_{11}+DG)|$, the nonlinear ODE (17) has stationary solutions at two distinct values of $\xi$. Each solution corresponds to an entrainment of the oscillator to the frequency of the incident field. Stability at the stationary point requires that we choose the solution with $\operatorname{Im}\{T_{12}\exp(-i\xi)/C(T_{11}+DG)\} < 0$. The phase at entrainment is complicated, but in the absence of detuning ($\omega = \Omega$), there is a simple expression for it:

$$\exp(-i\xi) = -i(DT_{12}/T_{22}(T_{11}+DG))^*/|DT_{12}/T_{22}(T_{11}+DG)| \qquad (18)$$

Comparison with 16 shows that the stable solution for $\xi$ corresponds to (16) having a positive right hand side, i.e., the stable phase at entrainment is such that the amplitude V of the oscillator is augmented relative to its value in the absence of an incident field. Increased V corresponds to increased nonlinear energy dissipation in the circuit. That conclusion is independent of the parameters T of the two-port network, and independent of the degree of detuning. Thus this would appear to be *stimulated absorption*, and the behaviors observed in the laboratory (see section IV) would appear to be not represented in this model.

Such a conclusion would be premature. While the flow of energy out of the nonlinear circuit into the transducer is, apparently, decreased by the incident acoustic field, the flow of energy out of the transducer into the medium differs from that by dissipation in the transducer. Thus the model may yet describe stimulated emission, but only if the transducer is dissipative and only if power dissipation within the transducer is lessened by the presence of the incident field.

Thus we are led to ask about the radiation of energy into the mechanical medium. The power radiated from the device is $\Pi = -\boldsymbol{f} \times d\boldsymbol{u}/dt = \boldsymbol{f} \times \{ -d\boldsymbol{u}^{inc}/dt + d\boldsymbol{G}/dt\ \boldsymbol{f}\ \}$ which is (by 14)

$$\begin{aligned}\Pi &= [T_{11}+DG]^{-1}(T_{21}v+Du^{inc}) \times dG/dt\, [T_{11}+DG]^{-1}(T_{21}v+Du^{inc}) \\ &\quad -[T_{11}+DG]^{-1}(T_{21}v+Du^{inc}) \times \{du^{inc}/dt\}\end{aligned} \qquad (19)$$

The terms in $v^2$ are almost identical to the spontaneous emission rate, differing only in that V is now slightly different. The terms in $\boldsymbol{u}^{inc\,2}$ are independent of V, and presumably due to passive losses on scattering off the dissipative parts of [T]. This presumption is supported by a short calculation that shows, for the case T = real, that the term in $u^{inc2}$ vanishes. The cross terms, in V x U, resemble stimulated emission and absorption. These terms are

$$\Pi_{UV} = |T_{11} + DG|^{-2}$$
$$[T_{21}v \times (dG/dt) Du^{inc} + Du^{inc} \times (dG/dt)T_{21}v - (T_{11} + DG)^* T_{21}v \times du^{inc}/dt] \quad (20)$$

On time averaging, we recover:

$$\Pi_{UV} = 2\Omega UV |T_{11} + DG|^{-2}$$
$$\text{Im}[T_{21}^* \exp(-i\xi)(T_{11} + DG^*)] \quad (21)$$

A substitution of the expression (18) for $\xi$ valid in the absence of de-tuning, gives

$$\Pi_{UV} = -2\Omega UV |T_{22}|/|T_{11} + DG||DT_{12}|$$
$$\text{Re}[T_{21}^{*2}(D/T_{22})^*[(T_{11} + DG^*)/(T_{11} + DG)^*] \quad (22)$$

If T is real, this is manifestly negative. [We recall that $D/T_{22} = C$ is the free capacitance and its real part is presumably positive.] We therefore conclude, as in the paragraph following eqn(18), that real T implies that the UV part of the power flow into the mechanical medium is negative, i.e, that the entrained oscillator absorbs energy from the field incident upon it.

If the elements of [ T ] are complex, the conclusion can differ. The expressions (21,22) are complicated, but permit some simplifications. For the case in which T is dominated by positive real parts of the elements on its diagonal, D and $T_{11}$ are very nearly real; the ratio $[(T_{11} + D G^*)/( T_{11} + D G)^*]$ is unity; $(D/T_{22}) \sim T_{11}$ is real and positive, and $\Pi_{UV}$ is positive if $T_{21}$ is imaginary *regardless of G*. Thus stimulated emission depends chiefly on the phase of $T_{12}$.

If we suppose instead that T is dominated by the real part of $T_{11}$, with all other elements being complex and of equal small order, then $D \sim T_{11} T_{22}$ and

$$\Pi_{UV} = -2\Omega UV |T_{22}|/|T_{11} + DG||DT_{12}|$$
$$\text{Re}[T_{21}^{*2} T_{11}(1 + T_{22}G^*)/(1 + T_{22}^* G^*)] \tag{23}$$

In this case the sign of $\Pi_{UV}$ depends on the phase of $T_{12}$ and also, if the magnitude of $T_{22}G$ is of order unity or larger, on the phase of $T_{22}$.

We conclude that at least for some sets of parameters the model exhibits stimulated emission. For sufficiently weak de-tuning, and for transducers with lossy electro-mechanical coupling, we expect an incident field to stimulate emission. A set of several such piezoelectric auto-oscillators on an acoustic cavity should behave such that their net energy emission scales with the square of the number of entrained oscillators, i.e, the system should be *superradiant,* [18] like some lasers. In the next section we illustrate this conclusion by means of numerical simulations.

IID  Amplitude dependent frequencies

In the above model, the natural frequencies of the auto-oscillators are independent of their amplitude. This is perhaps the simplest model, and it is encouraging that it permits stimulated emission. Nevertheless, we have seen signs of some amplitude dependence in our electronics, and it is not unreasonable to presume that, as its resistance varies with amplitude, so might the negistor's capacitance. Thus it may be indicated to investigate the implications of a generalizing the van-der Pol model to a Landau-Stuart form[9]. It may be that stimulated emission does not require that the transducers have lossy properties. It is outside our present purpose to investigate this further.

**III  Numerical Simulations**

The above theory may be illustrated by numerical simulation. We consider the following forced coupled nonlinear differential equation:

$$\ddot{\zeta} = -v/m$$
$$\dot{q} = I(v) + \dot{\zeta}$$
$$v - \eta I(v) = kq + \eta(\dot{\zeta} + \dot{u}) \tag{24}$$
$$\ddot{u} + \dot{u}/\tau + u = -\eta(\dot{u} + \dot{q}) + F_o \cos(\Omega t)$$

that describes a single limit cycle oscillator in contact with an acoustic cavity (here a single degree of freedom damped oscillator represented by $u(t)$) with an incident field due to a forcing $F_o$. We take the function $I(v) = \varepsilon(1-v^2)v$ (so $\beta$ is unity.) T and G for this system are

$$[\tilde{T}] = \begin{bmatrix} k + i\eta\omega & i\eta\omega \\ i\eta\omega & i\eta\omega \end{bmatrix}; \quad \tilde{G} = [1 - \omega^2 + i\tau\omega]^{-1} \tag{25}$$

This set of parameters may be thought of as describing the purely mechanical system pictured in figure 3.

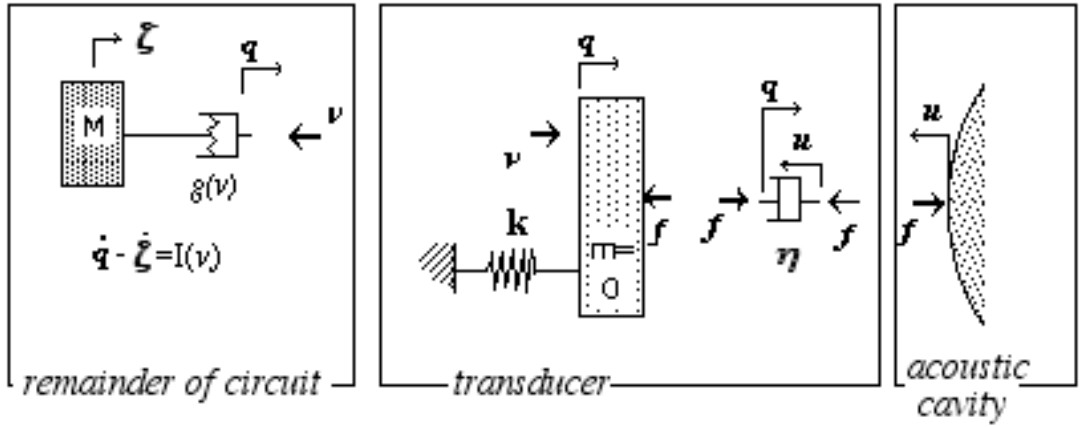

Figure 3] A purely mechanical version of our oscillator. The mass M plays the role of the inductor, the negistor is represented by the nonlinear damper whose differential velocity d($q$-$\zeta$)/dt is given by the function I($v$) of the force $v$ across it. Governing equations include force balance $v$-$f$=$kq$, and a constitutive relation for the linear resistance element $f = \eta$ d($q+u$)/dt.

We take $\Omega = k = 1$ without loss of generality, and choose $\sqrt{k/M}$ to be in the vicinity of $\Omega$. In order that the limit cycle be smooth, we take $\varepsilon$ to be small (typically, 0.1). To preserve the instability of the $q = u = 0$ solution in the absence of an incident field, we choose $\eta$ to be less than $\epsilon$; $\eta = 0.06$. This assures that the coupling is weak. So that G is smooth, we take $\tau = 1$. These choices lead to the following evaluations in the vicinity of $\omega = 1$

$$G = 1/[1-\omega^2+i\omega] \sim 2(1-\omega) - i$$

$$D = i\eta$$

$$C = T_{22}/D = 1 \qquad (26)$$

$$L = 1/M$$

$$T_{12} = T_{21} = T_{22} = i\eta\omega \sim i\eta$$

$$\omega = [1 - \epsilon^2/16] / \sqrt{M} + 5\eta^2/8; \quad \mu = \eta/2$$

The last line follows from theory that predicts the frequency of the limit cycle. The formula for the limit cycle frequency when not coupled to the acoustic cavity, $1/\sqrt{LC}$, was modified to account for finite $\epsilon$ [17], hence the term in $\epsilon^2/16$.

The coupled differential equations are solved by fourth order Runge-Kutta, where at each step the code must solve a cubic equation for $v$ in terms of $q$, $d\zeta/dt$ and $du/dt$. The code is started with generic nearly quiescent initial conditions, $q$, $\zeta$, $d\zeta/dt$ = random small << 1. Typical run times are thousands of cycles. All results are for the steady state.

In the case $F_0 = 0$, ie., without an incident field, theory above predicts values for the frequency of oscillation, for the amplitude V, and for the rate of spontaneous emission. Figures 4a,b show that the steady state frequency, for the case M =1 and various values of $\epsilon$ and $\eta$, indeed varies with $\eta$ like the prediction: $[1-\epsilon^2/16]/\sqrt{M} + 5\eta^2/8$.

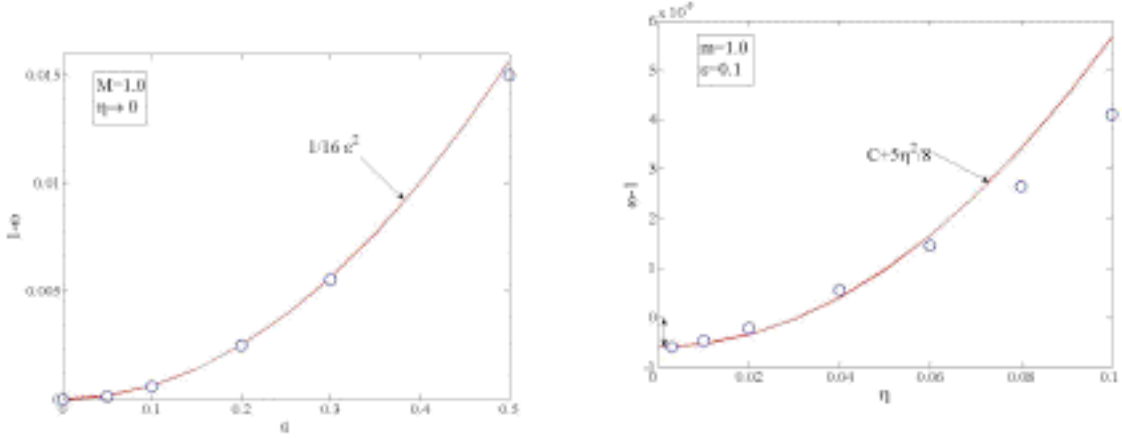

Figure 4] The dependence of the limit cycle frequency, obtained from numerical solution of Eqs. 26 at $F_0=0$ on $\varepsilon$ (a) and $\eta$ (b). The results of the simulations (symbols) are compared to the theoretical approximations (solid lines).

For the case of arbitrary $F_0$, we look for the presence of entrainment, such that the device, as represented by the time dependence of q, oscillates at the applied frequency $\Omega$ (=1) rather than the frequency it would take on its own. Theory above says that entrainment will occur for sufficiently weak de-tuning, $|\Omega-\omega| < (\eta/2) |U/V|$, where V is the amplitude of the autonomous auto-oscillator, approximately $[(\varepsilon\Omega-\mu)/6\varepsilon\Omega\beta]^{1/2} \sim 0.34$, and U is the amplitude of the incident field, $U = F_0/2$. For parameters $\eta=0.06$, $\varepsilon = 0.1$, entrainment is predicted for values of $|\omega-\Omega| < 0.045$ (at $F_0=1.0$). This is in good agreement with numerical simulation and with common understanding [13].

We also evaluate the average rate at which the device does work on the cavity

$$\Pi = -\eta < \dot{u} \, (\dot{u} + \dot{q}) >_{timeaverage}$$

and study in particular how this varies with $F_0$ and with detuning. This quantity is not discussed in the literature on entrainment[ - ], but as it is central to the proposed construction of a laser analog, it is studied here. Figure 5 shows the power inflow as a function of detuning for the case $F_0 = 0.5$ and $F_0 = 1.0$  It may be seen that the net inflow is positive in the limit of no detuning, and negative at high detuning.

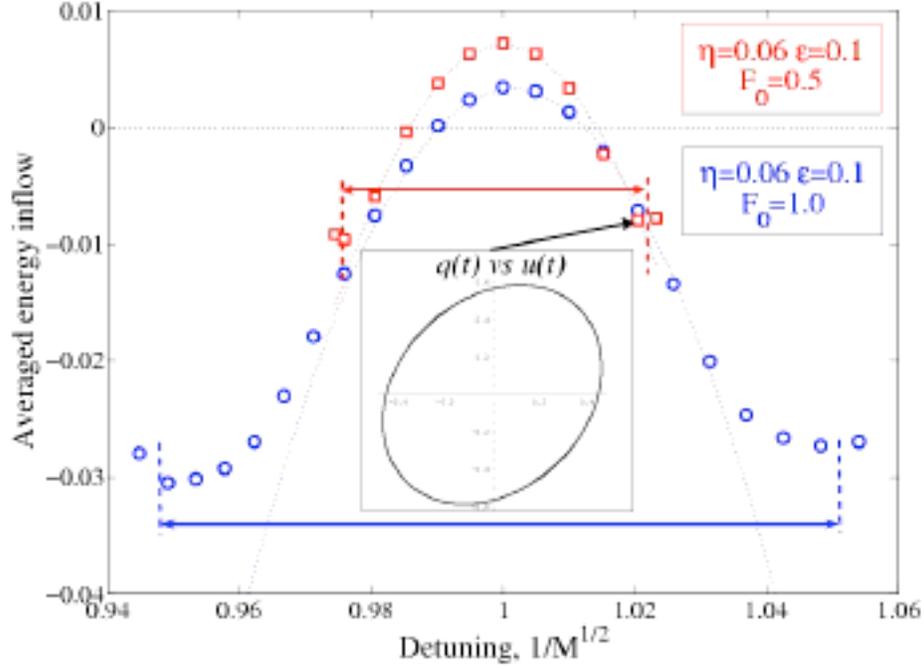

Figure 5] Entrainment (locking) of the oscillator to the driving incident field with $F_0=0.5$ (squares) and $F_0=1.0$ (circles).  The horizontal arrows indicate the regime over which the oscillator locks to the applied frequency.  The inset shows an example of the entrainment.

The character of the power flow is best illustrated, as in Fig 6a, as a continuous function of $F_0$, and for the case of no detuning.  At $F_o=0$, the emission is spontaneous and equal to 0.00093, which may be compared to the prediction of 0.0008 from eqn (13).   An analytic treatment of the specific model (24) predicts 0.00096.  At small $F_0$ the power inflow increases linearly with $F_0$, that is, with the incident wave amplitude $U = F_0/2$.  The slope of that dependence is predicted by the stimulated emission formula Eq. 22, which for the present parameters is $\Pi_{UV} = - 2\Omega U^{inc} V | T_{22}| / ( | T_{11}+DG| | DT_{12} | )$ **Re**$\{ T_{21}^{*2} (D/T_{22})^* [( T_{11} + D G^* )/ ( T_{11} + D G)^* ] \} = 0.017 F_o$, in rough accord with the parabolic fit's slope of 0.0235.  The difference is ascribed to changes in the terms $v^2$ neglected following equation (19). Figure 6a shows that the inflow is negative at large $F_0$, where the flow is dominated by passive losses proportional to $U^2$.

Figure 6b shows the predicted value $V = 0.034$ at $F_0= 0$, and the predicted (see discussion following eqn 18) increase of V for $F_0 > 0$. At $F_0^{cr}=1.15$ the total emission vanishes even at zero detuning, c.f. Figure 6a. At this point U, V and Q (the amplitude of *q*) all become equal to $F_0/2$. **v** and **q** oscillate in phase, and opposite to that of **u**. For $F_0$ greater then $F_0^{cr}$, the amplitude $U^{inc}$ exceeds V and the net flow is negative.

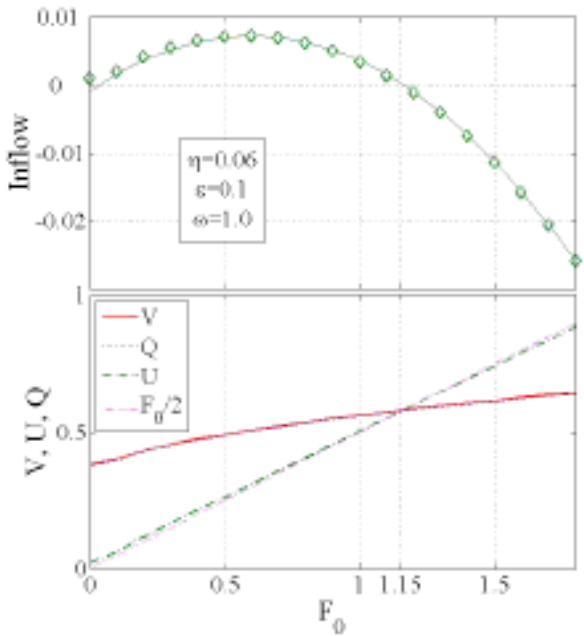

Figure 6] (a) Time-averaged emission/absorption rate as a function of $F_0$ (symbols) fitted with a parabola (solid line). The detuning is zero. (b) Limit cycle values of $U^{total}$, V and Q for different $F_0$. Crossing point U = Q at $F_0=1.15$ corresponds to the transition from net emission to net absorption.

Numerical simulations thus confirm key predictions of the theory above. In particular we see the familiar phenomena of spontaneous emission, and entrainment. We also see the predicted stimulated emission, and its quantitative accord with theory.

**IV Laboratory studies**

We have investigated these predictions in the laboratory by constructing the "negistor," or "lambda diode,"[19] illustrated in figure [7], and incorporating it into a LC oscillator where the capacitance is provided chiefly by a piezoelectric transducer.

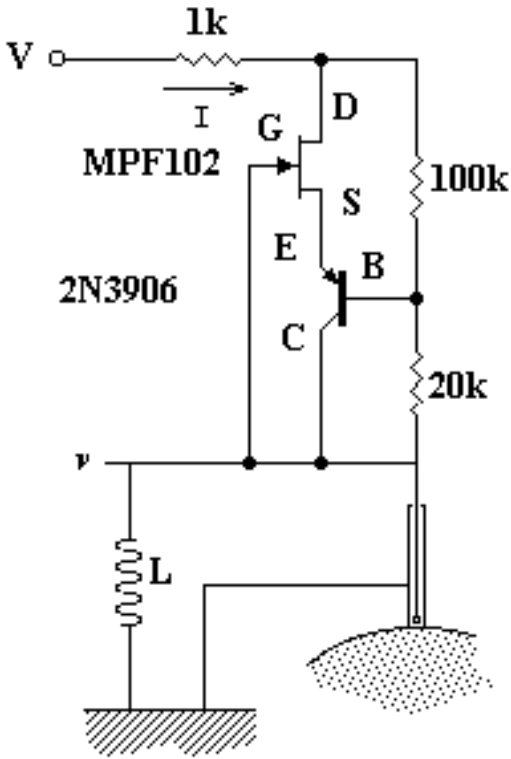

Figure 7] A lambda diode, or negistor, is constructed from two transistors. It has a current voltage relation as shown in the inset to figure 1. When it is incorporated into a circuit with an inductor L ( ~ 220 µH) and a capacitance C as provided by the piezoelectric transducer and its cables (~200 pf), the circuit auto-oscillates at a frequency of about $\omega = 1/\sqrt{LC}$.

We find that the circuit auto-oscillates in a periodic nearly harmonic limit cycle whose frequency is tunable by varying L or by adding additional capacitance. Here, and in all cases below, the spectral width of the auto-oscillation is finer than our precision of 1 to 10 Hz. Line width appears to be governed by background noise and spectral intensity as is the Schawlow-Townes line width of a laser. It has recently been investigated for an ultrasonic system with gain [20]. When the piezoelectric transducer is then attached to an elastic body, the frequency of oscillation shifts, as in figure [8]. Frequency changes can be different depending on the position of attachment, the material, the size of the solid body, or the presence of oil couplant. Frequencies always increase when the transducer is placed on the body, reflecting an increase in effective stiffness. No change in frequency is attendant upon additional grounding of the transducer case, so changes may be ascribed to the mechanics, not the electronics. Changes are small compared to natural frequency of the order of 500 kHz (as specified by the choice of L and additional capacitances).

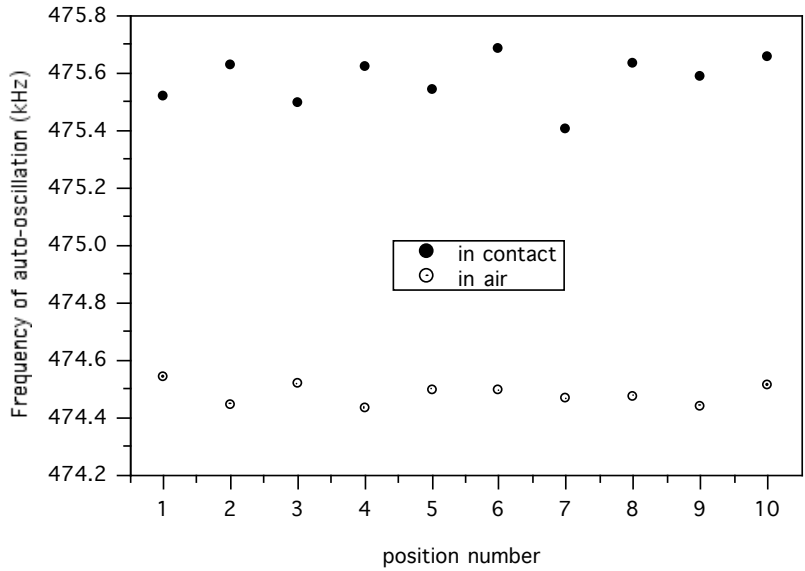

Figure 8] The frequency of auto oscillation varies by about 1 kHz as the transducer is alternately in contact (filled circles) and out of contact (open circles) with the aluminum block. Slight variations in the frequency of the non-contact case are ascribed to stray capacitances due to the operator's hands. Variations in the attached case are ascribed to variations in coupling strength as the transducer is re-attached, and perhaps to variations in local Greens function.

Theoretical frequency$^2$ is given by the real part of $[\ (LC)^{-1} + \{\ T_{12}\ T_{21}\ \Omega^2\ /\ T_{22}\ [\ T_{11} + DG]\ \}\ ]$. From the observed magnitude of the frequency changes in figure 8 as G is alternated between ∞ and the G of the elastic body in different places, and based upon statistically identical data from steel where G is smaller than it is in aluminum, we estimate Re $(T_{12}\ T_{21}\ /\ T_{22}\ T_{11})$ to be of the order 0.004, and DG to be less than $T_{11}$. It is worth noting that in a system with high modal overlap (meaning level-spacings are much less than absorption widths) such as the one used in generating figure 8, $G_{xx}$ has fluctuations, from place to place or frequency to frequency, that are small;

$|\delta G_{xx}|/|G_{xx}| \ll 1$ [21]. G is very nearly equal to the value it would take in an infinite half space. For this reason the frequency of auto-oscillation ought to depend only weakly on position *x*.

We study the occurrence of entrainment by adding an incident field by means of the transducer at point *z* in figure 1. We seek the conditions under which the nonlinear electronic circuit adopts the frequency of the applied field. Figure [9] shows (continuous curve) a short section of the spectrum of the transfer function between the generator and the monitor, $|h_{zy}|$ on a 70 mm aluminum cube with a surface treatment that enhanced dissipation. ( On ignoring scattering by the transducers one can calculate this transfer function in terms of other quantities that have been defined here: $h_{zy} \approx T_{12} G_{zy} T_{21} / (1+T_{22}G_{zz})(1+T_{22}G_{yy}) \sim T_{12}^2 G_{zy}$.) The bold vertical lines indicate the frequency of the generator at *z* (757.8kHz) and that of the undisturbed oscillator in contact with the solid (758.63kHz). The several isolated points indicate the frequency, and rms amplitude, received at *y*, due to the oscillator, at each of eleven equally spaced generator amplitudes from 0.0 to 5.0 volts. It may be seen that the oscillator frequency is pulled towards that of the generator, and that the degree of pulling is monotonic in generator strength. This behavior is familiar from the literature on entrainment of auto-oscillators. Less familiar are the occasional discontinuities. The two discontinuities, between 3.5 and 4.0 volts, and between 4.5 and 5.0 volts where locking ensues, appear to correspond to features in the transfer function h of the passive cavity. They are related to the random reverberant nature of the wave propagation and are not found in the standard model of frequency-independent coupling.

Figure 10 shows the behavior of the entrainment as the frequency of the applied cw signal from the generator is varied, at a fixed amplitude of 5.0 volts. Frequency is varied in 0.2 kHz steps from 757.0 to 759.8 kHz. The natural frequency of the auto oscillator in contact with the body is indicated by the thin vertical line at 758.63 kHz. In the absence of full entrainment the auto-oscillator frequency is pulled towards that of the generator. It may be seen that the auto-oscillator is entrained to the applied frequency if the applied frequency is close to the natural frequency, and if the signal from the generator at *z* as received at the auto-oscillator at *x* is strong enough, ie., if the transfer function is large. Entrainment is thus a complicated non-monotonic function of applied frequency.

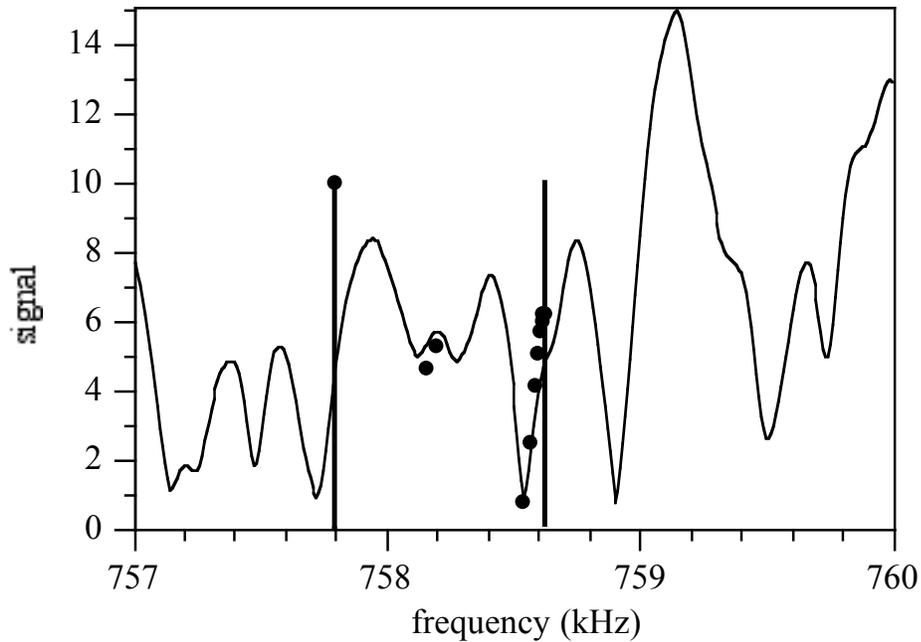

Figure 9] The nonlinear oscillator's frequency (isolated points) is pulled, monotonically and discontinuously, as the signal strength from the generator at *y* is increased from 0 to 5 volts in 0.5 volt steps. The continuous curve is the transfer function h between monitor and generator. The bold vertical lines mark the positions of the spectral lines (one at 757.79 due to the generator, another at 758.63 due to the uninfluenced auto-oscillator) at vanishing voltage from the generator.

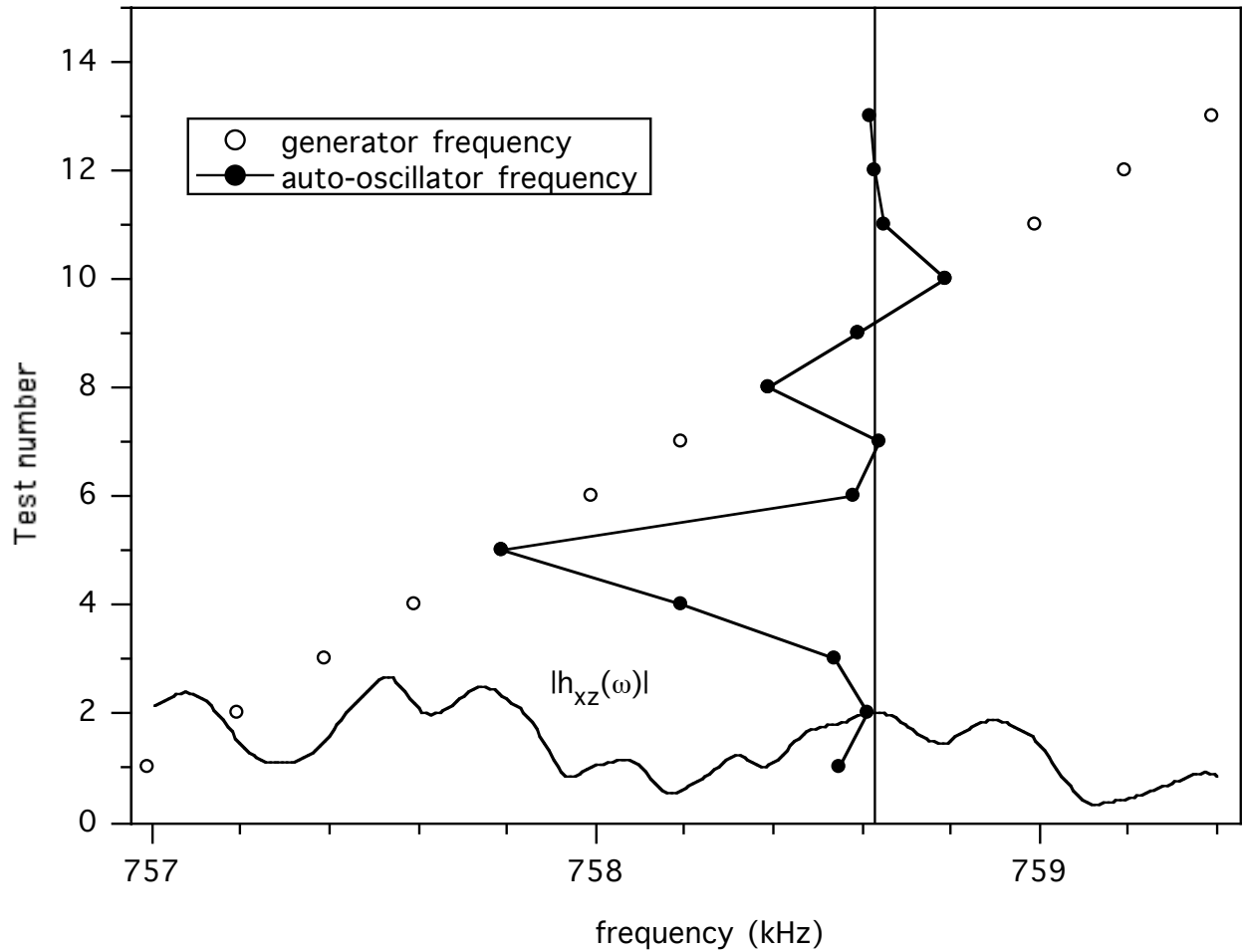

Figure 10] A study of entrainment versus the frequency of the entraining field at fixed cw source amplitude of 5 volts. The frequency of the auto-oscillator at $x$ is 759.63 kHz in the absence of an applied field, as indicated by the vertical line. The irregular smooth curve shows the (scaled to fit) transfer function $|h_{xz}(\omega)|$ between the source of the entraining field and the position of the auto-oscillator. Open circles indicate the frequency of the entraining field as specified by the operator. Closed circles indicate the frequency of the auto-oscillator as it is pulled towards that of the entraining field. The structure of the entrainment is complex, and the process is non-monotonic and discontinuous, influenced as it is by the complicated function $h(\omega)$.

We also studied the energy in the system as a function of the amplitude of the applied field, and did so with primary focus on the case of no de-tuning, i.e, for $\Omega = \omega$. Being unable to measure the power inflow at *x*, we instead measured the signal strength at *y* and interpreted it as a measure of the (square root of the) energy in the cavity. This is an imperfect measure of acoustic energy unless the system has low modal overlap and the frequency of interest is near one of the modes, so that only one natural mode is excited. Thus we chose a cavity that is a thin rod, with low modal density and modest absorption, as pictured in figure 11.

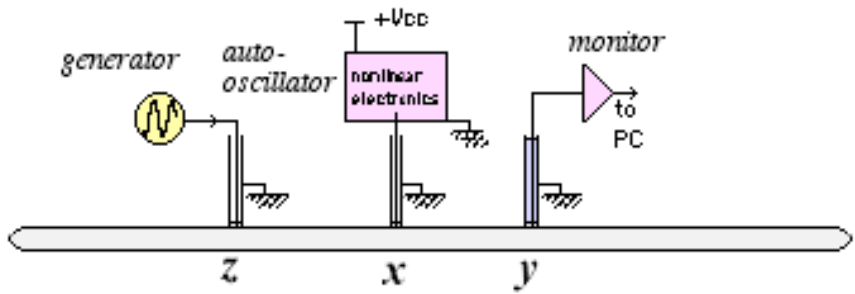

Figure 11] The system is attached to an aluminum rod of 15 cm length and 3.16 mm diameter. Below 580 kHz, it has four propagating guided modes of elastic waves.

We also recognize that the total energy in the cavity is proportional to the sum of the power flows from the nonlinear circuit at *x* and the generator at *z*. The energy generated in the acoustic cavity may be decomposed in two terms, each equal to the time-average of the dynamic force at a transducer times the material velocity at the same point. This quantity at *x*, it was shown above, has a term in $V^2$ and a term in UV, where U is the field incident from the prescribed forcing at *z* and proportional to the cw signal input strength 'g' from the generator. The work done at *z*, however, includes not only a term in $g^2$, but also a term in gV due to the field incident from *x*. Thus the energy produced in the acoustic cavity includes the intuitive $g^2$ and $V^2$ terms, and the gV term described above, but another term scaling like gV that was not discussed above. Analysis shows that the excess stimulated emission of energy at *z* is of the same order as that at *x*, but with a phase depending on the phase of $G_{gu}$. It is not necessarily positive, even if $\Pi_{UV}$ is positive. Nevertheless, for the case of low modal overlap, the two terms have the same sign and the total energy in the cavity can be a proxy for the power flow from the nonlinear circuit.

Figure 12 shows the rms of the cw signal observed at the monitor at *y* for various levels of the applied cw input at *z*.  Stimulated emission is apparent in the positive slope of the upper curve at g=0.  Figure 12b shows the difference of the squares of the curves in 12a, thus corresponding to the energy of spontaneous emission at g=0 plus the UV part linear in g that corresponds to stimulated emission.   Stimulated emission is apparent in the positive slope.  Power output is greater than the sum of the power outputs from the nonlinear circuit and the generator when they operate alone.

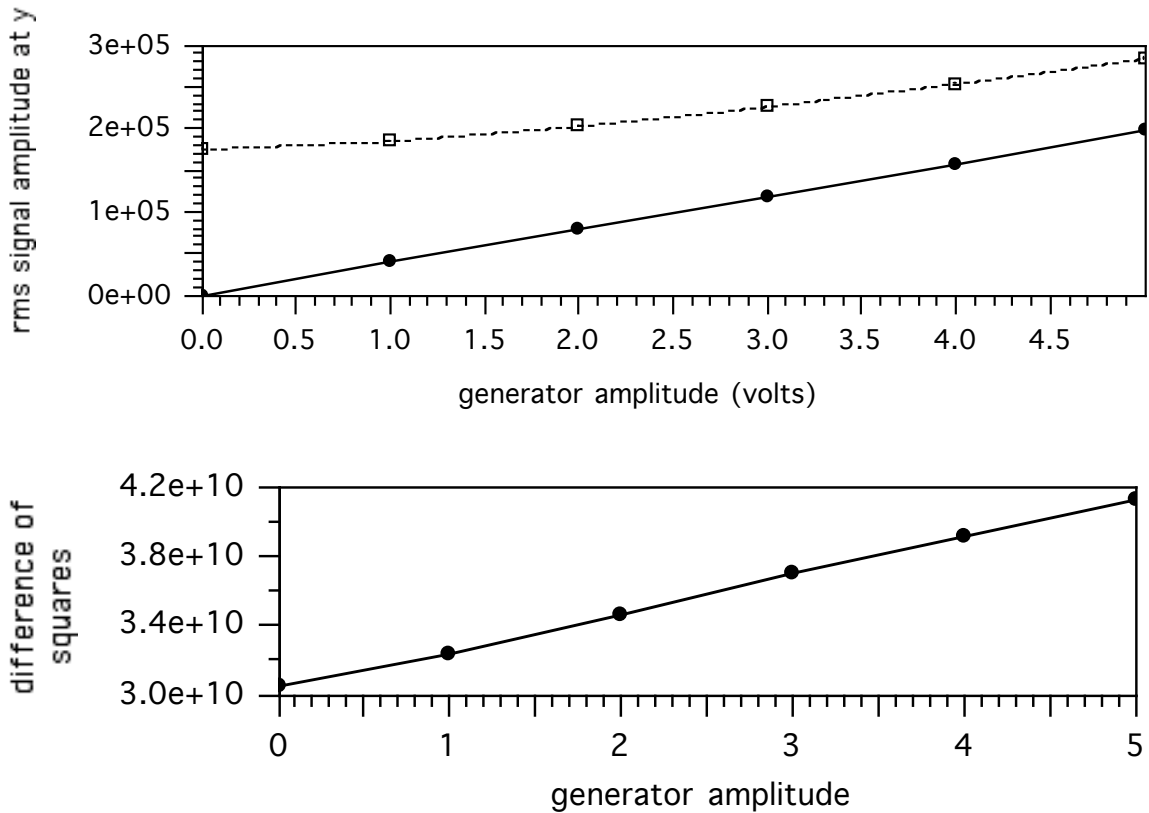

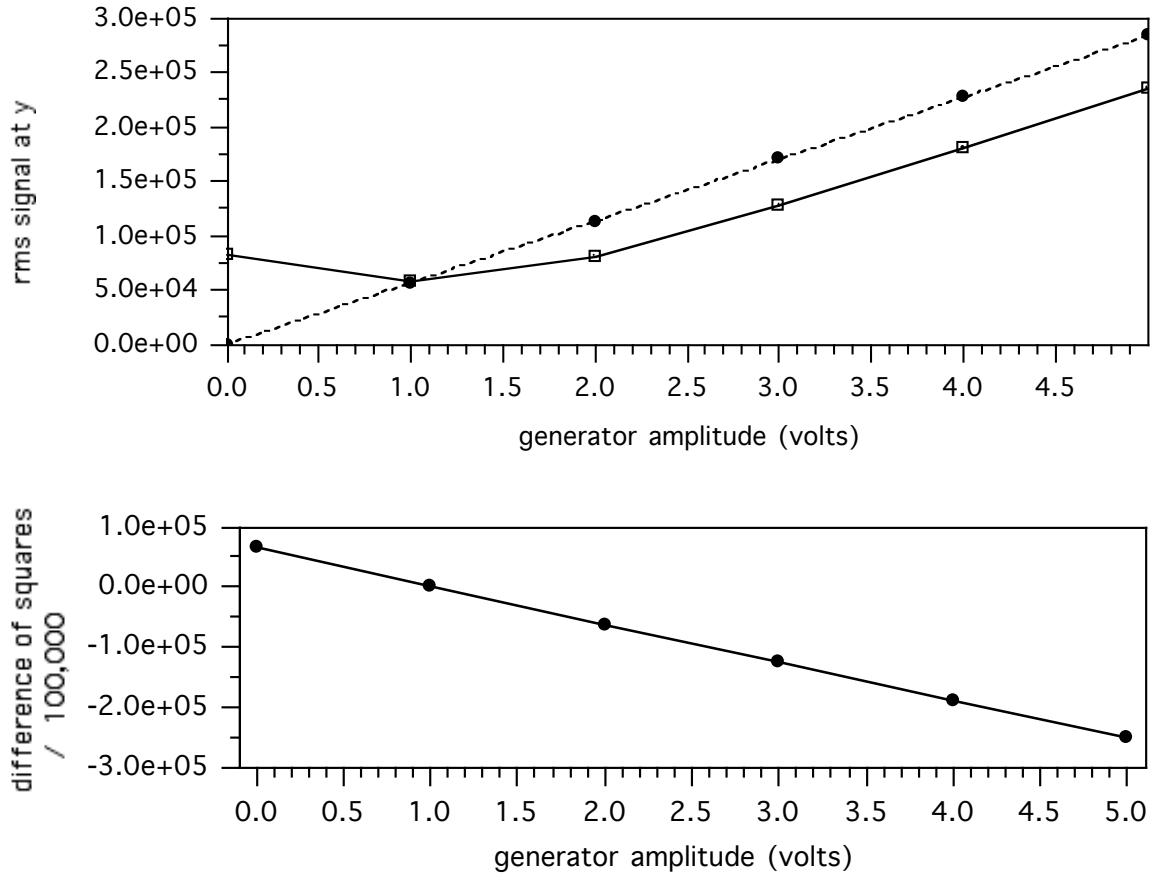

Figure 12] Evidence of stimulated emission and absorption in the rod of figure 11. The rms signal amplitude at y is plotted against the amplitude of the signal from the generator into z, for the case of the auto oscillator powered (filled circles) and unpowered (empty squares). The difference of their squares is also plotted, indicating a coherent interference, i.e stimulated emission (top case) and absorption (bottom case).

**V Uasing**

On replacing the generator–driven transducer at z with one or more additional van der Pol auto-oscillating transducer circuits, as in figure 13, we have what may be termed a "uaser" (Ultrasound amplification by stimulated emission of radiation [7]), an acoustic analog to a laser. The system is similar to that encountered when two or more Josephson junctions couple and synchronize through their microwave radiation field [8] also noted as analogous to a laser. The data presented here are not meant to fully elucidate the properties of such systems. Rather these illustrations are presented merely to provoke imagination and suggest further studies. A thorough study awaits development of a theory to inform such experiments.

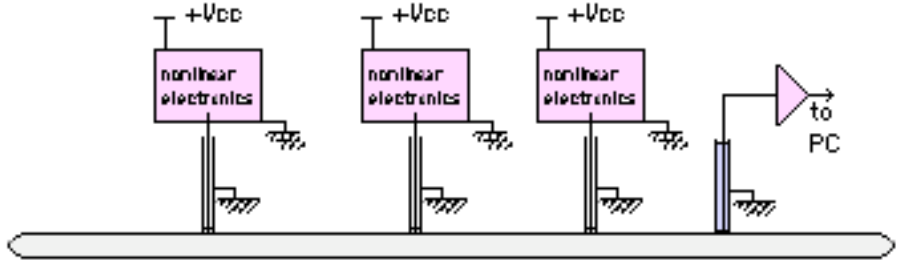

Figure 13] Two or three autonomous nonlinear piezoelectric oscillators are placed in contact with the aluminum rod of figure 11. The acoustic state is monitored with a passive detector.

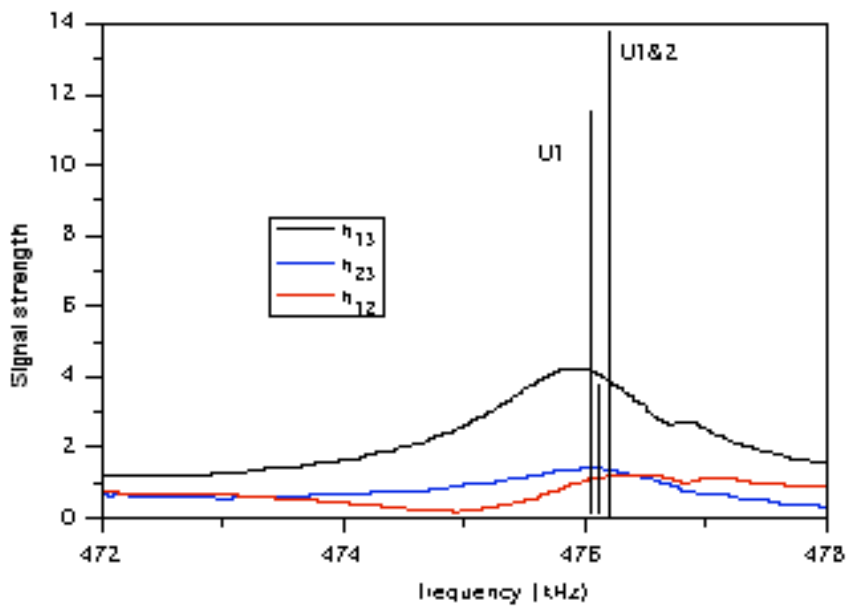

Figure 14a] Two auto-oscillators are placed on the rod of figure 13 ( at positions 1 and 2). The resulting signal is monitored at position 3. Three transfer functions h (continuous lines) are plotted. We also plot the rms amplitude and frequency at the monitor (bold vertical lines) for each of three cases: only auto-oscillator number 1 powered (label U1), only oscillator number 2 powered (U2, not labeled in figure), and both powered (label U1&2). On normalizing these amplitudes to the strength of the transfer functions, we find that U1&2 is within 2% of the sum of U1 and U2. The two auto-oscillators not only lock to a single frequency, but contribute in phase to the signal at the monitor.

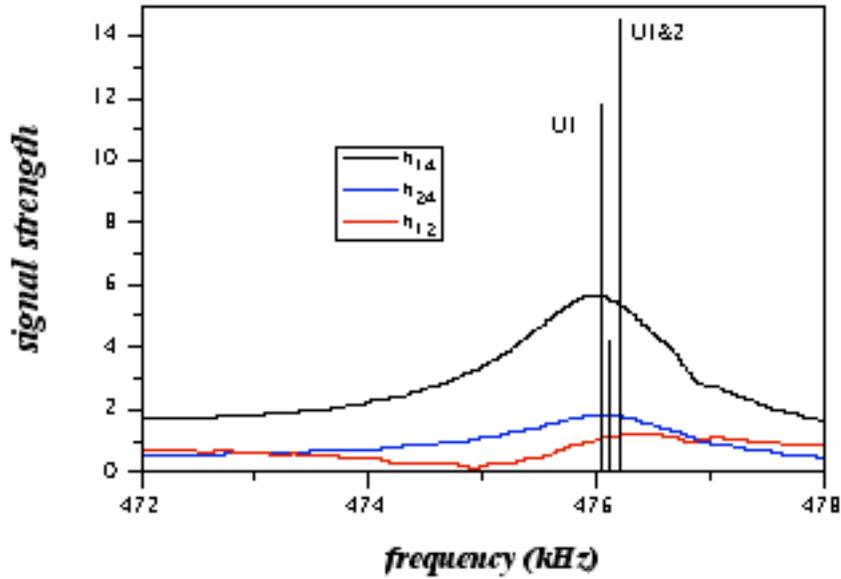

Figure 14b] as in 14a, but with a different position for the monitor (4). Again the amplitudes add.

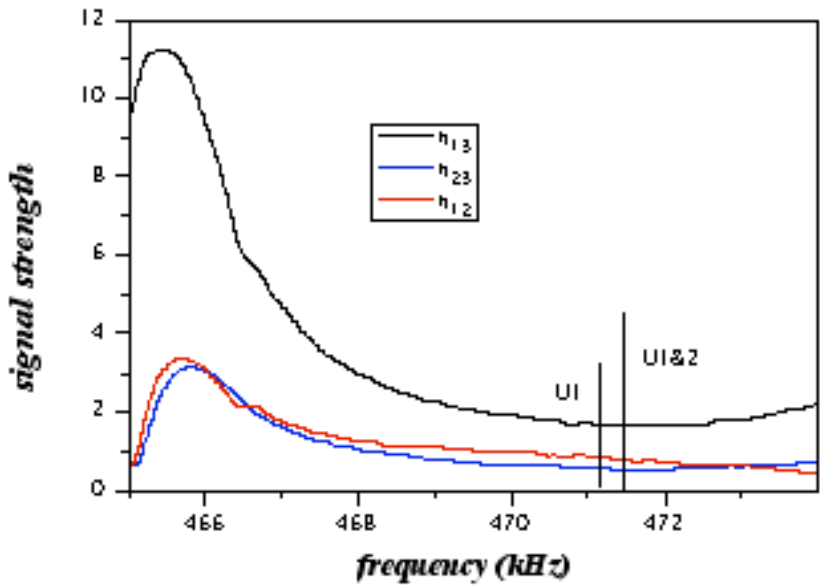

Figure 14c] As in 14a, but with the auto-oscillators tuned to a different frequency range. Again the amplitudes add to within 2%. Not shown, a figure analogous to 14b with monitor at position 4, but for the case of tuning near 471.5 kHz, and for which the amplitudes again add.

Figures 14 compare the rms amplitude detected by the monitor, and the frequency of that signal, on powering one, or two, auto-oscillators. 14a) and 14b) are for a case in which both auto-oscillators are tuned to a frequency near 476 kHz, near a maximum in the transfer functions. Regardless of the position (3 or 4) of the monitor, we find that the (normalized) amplitudes add; that is, the net energy in the cavity is greater than the sum of the energies due to each oscillator alone. Normalization is effected by replacing the rms amplitude U1 at frequency $f_1$ with U1 $h_{13}(f_{12})/h_{13}(f_1)$, and U2 with U2 $h_{23}(f_{12})/h_{23}(f_2)$. We then compare U12 with U1 $h_{13}(f_{12})/h_{13}(f_1)$, + U2 $h_{23}(f_{12})/h_{23}(f_2)$. 14c) shows the same phenomenon when the oscillators are tuned to a frequency near 471 kHz, a minimum in the transfer function. For almost all cases we have investigated on the rod, the amplitudes add. Occasionally, they subtract. It is apparent that this synchronization is similar to that described elsewhere[ 9-12, esp 13 chapters 10 and 11] except that here we also emphasize issues of wave power generation, stimulated emission and super radiance.

Figure 15] shows a case of three auto-oscillators. For reference, we plot the three transfer functions $h_{14}$, $h_{24}$, and $h_{34}$ between the auto-oscillators and the monitor at position 4. Bold vertical lines indicate the (un-normalized) rms amplitudes of each individual auto-oscillator, of each pair, and of all three. The normalized amplitudes of individual oscillators 1 and 2 (6.53 and 2.43 when referred to frequency 476.114) add to approximate the amplitude of that pair together (8.27). The normalized amplitudes of oscillators 2 and 3 (2.45 and 7.15 when referred to frequency 476.095) add to approximate the amplitude of that pair together (9.20). These pairs both add constructively. Oscillators 1 and 3 have normalized amplitudes (5.19 and 4.73 when referred to frequency 476.482) that roughly subtract to approximate the amplitude (1.05) of that pair together. Oscillators 1 2 and 3 (normalized amplitudes 7.31,

2.66, and 7.88 respectively when referred to frequency 475.959) very nearly add to approximate the amplitude of all three together (15.93). Acoustic energy is generated coherently, and at more than twice the rate that would be obtained in the absence of feedback. The system is super-radiant.

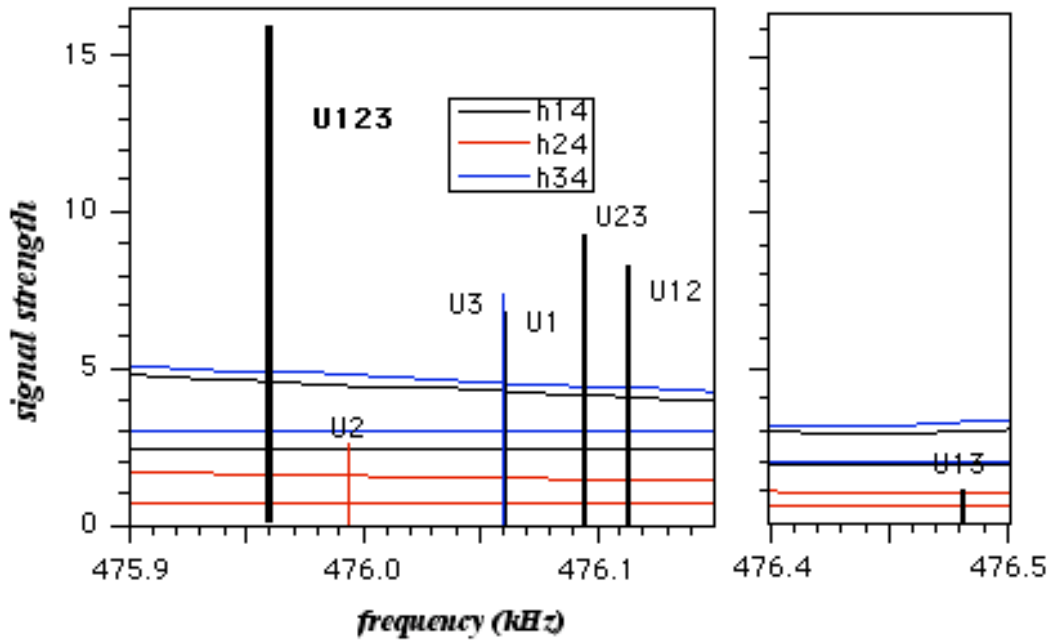

Figure 15] Three auto-oscillators are monitored by a single receiver at point 4. The rms signal strengths (rescaled to fit on the plot) are indicated by the vertical lines. Amplitudes of single auto-oscillators (thin vertical lines) are compared to those when auto-oscillators are taken in pairs (wider lines), and when all three auto-oscillators are powered (widest line).

**VI Summary**

It has been shown that nonlinear van-der Pol-like piezoelectric oscillators can be configured to exhibit the key behaviors required of sets of neighboring continuously pumped "atoms" in a classical analog for a laser. These include frequency locking, stimulated and spontaneous emission, stimulated absorption, and super-radiance. We conjecture that the principles illustrated here will find direct application in the construction of new kinds of acoustic generators, and indirect application in scale-model emulation of laser dynamics, in particular in research on random[22] and chaotic[23] and photonic crystal lasers[24].

**Acknowledgments** This work was supported by the NSF CMS 05-28096. AY acknowledges support from University of Missouri-Rolla.